ORIGINAL ARTICLE

# LoDAvatar: hierarchical embedding and selective detail enhancement for adaptive levels of detail Gaussian avatars

Xiaonuo Dongye[1] · Hanzhi Guo[1] · Le Luo[2] · Haiyan Jiang[3] · Yihua Bao[1,4] · Jie Guo[2] · Zeyu Tian[1] · Dongdong Weng[1,5]

**Abstract**
With the advancement of virtual reality, the demand for 3D human avatars is increasing. The emergence of Gaussian Splatting technology has enabled the rendering of Gaussian avatars with superior visual quality and reduced computational costs. Despite numerous methods researchers propose for implementing drivable Gaussian avatars, limited attention has been given to balancing visual quality and computational costs. In this paper, we introduce LoDAvatar, a method that introduces levels of detail into Gaussian avatars through hierarchical embedding and selective detail enhancement methods. The key steps of LoDAvatar encompass data preparation, Gaussian embedding, Gaussian optimization, and selective detail enhancement. We conducted experiments involving Gaussian avatars at various detail levels, employing objective assessments and subjective evaluations. The outcomes indicate that incorporating levels of detail into Gaussian avatars can decrease computational costs during rendering while upholding commendable visual quality, thereby enhancing runtime frame rates. We advocate adopting LoDAvatar to render multiple dynamic Gaussian avatars or extensive Gaussian scenes, balancing visual quality and computational costs.

## 1 Introduction

The progression of virtual reality (VR) technologies has increased the demand for realistic 3D human avatars (Zheng et al. 2022). Traditional methods of creating 3D human avatars often involve utilizing scan data or performing 3D modeling based on multi-view images, resulting in mesh avatars (Jiang et al. 2023). In recent years, the introduction of 3D Gaussian Splatting (3DGS) (Kerbl et al. 2023) technology has opened up new avenues for generating human avatars. Gaussian Splatting is an innovative rendering technique designed for real-time rendering of virtual objects and scenes. In contrast to conventional methods that rely on points and meshes for virtual object and scene

✉ Le Luo
leluo1989@gmail.com

✉ Dongdong Weng
crgj@bit.edu.cn

Xiaonuo Dongye
dyxn@bit.edu.cn

Hanzhi Guo
hanzhiguo@bit.edu.cn

Haiyan Jiang
haiyan.jiang@polyu.edu.hk

Yihua Bao
boye1900@outlook.com

Jie Guo
guoj01@pcl.ac.cn

Zeyu Tian
tianty97@163.com

1 Beijing Engineering Research Center of Mixed Reality and Advanced Display, Beijing Institute of Technology, No.5 Yard Zhongguancun South Street, Beijing 100081, China

2 Pengcheng Laboratory, No.6001 Shahe West Road, Shenzhen 518055, Guangdong, China

3 Research Center for Art and Culture Technology and PolyU NVIDIA Joint Research Centre, The Hong Kong Polytechnic University, No.11 Yuk Choi Road, Hksar 999077, China

4 SoulShell(Beijing) Technology Co., Ltd, No.17 Guangshun North Street, Beijing 100102, China

5 Zhengzhou Academy of Intelligent Technology, Beijing Institute of Technology, No.12 Qidizhengzhou Science City, Zhengzhou 450046, Henan, China construction, 3DGS is presented as a flexible and expressive representation (Fei et al. 2024). Anisotropic 3D Gaussians can accurately depict high-quality radiation fields, and these Gaussians are explicit and well-suited for GPU-based rapid rasterization (Tang et al. 2023). This capability enables rendering high-quality virtual avatars in VR while reducing computational costs and achieving high frame rates during rendering (Chen and Wang 2024). In creating dynamic Gaussian avatars, researchers have delved into the methodologies of drivable 3D Gaussian Splatting (Chen and Wang 2024). Drivable 3D Gaussian Splatting embeds Gaussians onto the surface of the corresponding mesh avatar, whereby transforming Gaussians from the world coordinate system to the local coordinate system on the surface of the corresponding mesh triangle (Yang et al. 2024). This allows the Gaussians to change with the mesh model, enabling the dynamic rendering of Gaussian avatars. Due to its reduced data storage requirements and the ability to perform actions beyond the captured data set, drivable 3DGS is widely used in dynamic Gaussian avatars (Zielonka et al. 2023). This method can generate Gaussian avatars based on existing mesh avatars and has been widely used in previous research, such as Gaussian Avatars (Qian et al. 2024) and Splatting Avatars (Shao et al. 2024).

Although research has focused on creating dynamic Gaussian avatars, little attention has been paid to balancing the visual quality and computational costs. Increasing the number of Gaussian avatar generations can increase visual quality but simultaneously increase computational costs (Liu et al. 2024). Real-time virtual avatar interaction in VR requires efficient rendering to achieve high frame rates. Our motivation is utilizing a manageable number of Gaussians to generate avatars and introduce levels of detail (LoD) on Gaussian avatars to better leverage the advantages of high visual quality and low computational costs inherent in Gaussian Splatting.

To balance superior visual quality and minimize computational costs, this paper introduces LoDAvatar, which generates Gaussian avatars with varying LoD through hierarchical embedding and selective detail enhancement methods, as depicted in Fig. 1. The methodology comprises four phases: data preparation, Gaussian embedding, Gaussian optimization, and selective detail enhancement. In the data preparation phase, we utilize existing mesh avatars or captured multi-view key frame images as input to generate the homogeneous topological mesh avatar and a sequence of key frame images. Additionally, we record the multi-view key frame images along with their corresponding camera parameters. Subsequently, the Gaussian embedding involves establishing a local coordinate system on each triangle face of the mesh avatar. Gaussians are initialized at each triangle's vertices and surface centers, with their parameters transformed from the world coordinate system to the local coordinate system. Following Gaussian embedding, we conduct Gaussian optimization by constraining the Gaussian positions at the triangle vertices and maintaining a constant number of Gaussians. After optimization, the Gaussians at the original triangle centers are repositioned and connected to the positions of the Gaussians of the vertices to form new triangles. In the selective detail enhancement phase, image masks are employed to identify triangles that require detail enhancement. New Gaussians are embedded on the corresponding triangle faces to achieve selective detail enhancement. Subsequent optimization procedures maintain the low-detail Gaussians fixed while refining solely the newly introduced Gaussians in each iteration, progressively enhancing the avatar's details and generating Gaussian avatars spanning from low to high levels of detail. Compared with prior work, LoDAvatar achieves

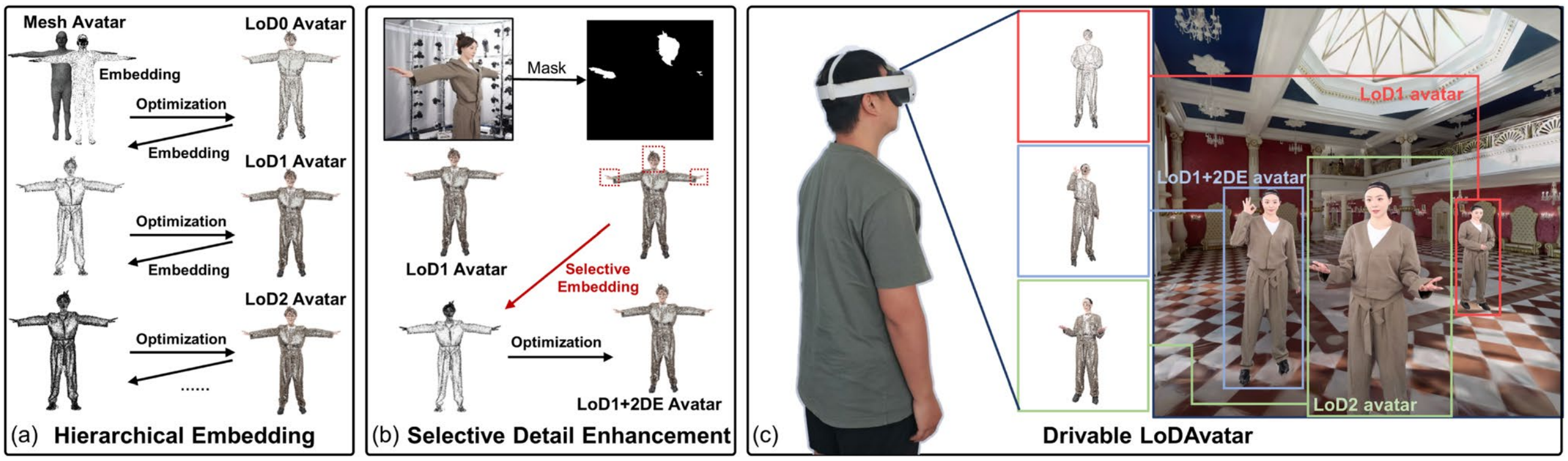

**Fig. 1** The LoDAvatar introduces levels of detail on Gaussian avatars, achieving a balance between visual quality and computational costs. The implementation of LoDAvatar relies on hierarchical embedding and selective detail enhancement methods. **a** The hierarchical embedding method starts from a base mesh avatar and progressively generates Gaussian avatars, ranging from lower to higher LoD. **b** Selective detail enhancement allows detail to be enhanced only at specific areas using image masks, allowing precise control of the number of Gaussians, thereby reducing computational costs. **c** Avatars generated through LoDAvatar can be driven in real-time and are well-suited for integration into VR

more precise control over the number of Gaussians through Gaussian embedding, optimization, and selective detail enhancement, enabling the generation of Gaussian avatars at various levels of detail. These avatars, which can be driven and used for VR interactions, demonstrate superior visual quality and reduced computational costs.

In summary, our main contributions are threefold:

1. We introduce a novel method for hierarchical embedding within mesh avatars, facilitating the drivable Gaussian avatars. This method enables the generation of human avatars with varying levels of detail by regulating the number of Gaussians, thereby striking a balance between visual quality and computational costs.
2. We use selective detail enhancement for area-controllable levels of detail by incorporating the image mask in conjunction with hierarchical embedding. By enhancing the avatar's facial and manual features, we imbue the avatar with enhanced details, rendering it more suitable for real-time interactions within VR.
3. We evaluated our proposed method through objective and subjective evaluations to showcase the high visual quality and reduced computational costs.

## 2 Related work

### 2.1 Gaussian splatting for virtual 3D construction

The recent emergence of 3D Gaussian Splatting technology has introduced a novel approach to generating high-quality virtual objects and virtual avatars (Dalal et al. 2024). This technology utilizes multi-view images to initialize a Structure from Motion (SfM) point cloud, which is subsequently transformed into 3D Gaussians. These points are projected onto the image plane using aligned cameras and rendered through differentiable rasterization. Following this, the rendered image is compared to the input multi-view images to compute loss, update the parameters within the 3D Gaussians, and adjust the Gaussians through adaptive density control. We summarize the notable features of Gaussian Splatting as follows: (1) The image-based 3D reconstruction approach expedites rapid data acquisition and virtual object generation. (2) It ensures the visual quality of virtual objects. (3) It incurs lower computational costs.

Gaussian Splatting technology demonstrates strong performance in digital avatar creation (Hu et al. 2024; Liu et al. 2024), real-time rendering (Lu et al. 2024; Jiang et al. 2024; Gao et al. 2023), and scene generation (Katsumata et al. 2023; Li et al. 2023; Yan et al. 2024). In avatar creation, researchers have focused on rendering dynamic face (Wang et al. 2023) and full-body avatars (Pan et al. 2024). For instance, Saito et al. introduce relightable Gaussian codec avatars to create high-fidelity relightable head avatars with animated expressions (Saito et al. 2024). Zheng et al. present GPS-Gaussian, a method for synthesizing novel views of characters in real time (Zheng et al. 2024). Given the impressive attributes, 3DGS has emerged as a significant method for generating virtual avatars in VR (Chen and Wang 2024).

### 2.2 Gaussian embedding and dynamic Gaussian avatars

While 3DGS has demonstrated exceptional rendering capabilities for static objects and environments, integrating dynamic objects and avatars into static scenes presents significant challenges (Lee et al. 2024). Researchers have explored two primary approaches for dynamic Gaussians, i.e. 4D Gaussian Splatting and drivable 3DGS (Chen and Wang 2024).

The 4D Gaussian Splatting introduces a temporal dimension to the 3DGS, capturing Gaussian objects at each frame through 4D data acquisition and rendering them frame by frame (Yang et al. 2023). A notable example of this approach is HiFi4G (Jiang et al. 2024), which combines 3D Gaussian representation with non-rigid tracking to achieve a compact and compression-friendly representation.

Drivable 3DGS involves dynamically adjusting the parameters of individual Gaussians within the Gaussian objects during rendering to achieve dynamic effects (Yang et al. 2024). This approach, requiring less data storage, has found wide application in dynamic scenes (Luiten et al. 2023), facial expressions (Ma et al. 2024), hair rendering (Luo et al. 2024), body avatars (Huang et al. 2024; Kocabas et al. 2024), etc. Gaussian embedding is a crucial method in the implementation of drivable 3DGS. The fundamental concept of Gaussian embedding involves rigging the Gaussian representation to a parametrically deformable mesh (Hu et al. 2024). As the mesh deforms, the embedded Gaussians on its surface adjust accordingly. For instance, Qian et al. introduced Gaussian Avatars (Qian et al. 2024), which utilize Gaussian initialization at the surface center of each mesh to create a series of dynamically bound Gaussians through adaptive density control. Shao et al. propose Splatting Avatars (Shao et al. 2024), employing trainable Gaussian embedding on a standard mesh. This method generates a set of Gaussians randomly on the mesh surface and refines them using the walking on triangles (Shao et al. 2024) technique. Furthermore, Jiang et al. developed the VR-GS system (Jiang et al. 2024), a highly efficient two-level embedding strategy that enables the interactive representation of physical dynamics-aware Gaussians in VR. Despite the various Gaussian embedding methods researchers have proposed, the challenge of efficiently implementing Gaussian embedding remains unresolved (Lee et al. 2024). While the studies above have focused on achieving dynamic Gaussian avatars, the issue of controlling the number of Gaussians in dynamic Gaussian avatars remains an open research question. Our research, inspired by the principles of Gaussian embedding, aims to devise low computational loss Gaussian embedding.

### 2.3 Levels of detail

Levels of detail play a crucial role in managing the intricacy of virtual environments to strike a balance between complexity and performance in the realm of computer graphics (Luebke et al. 2002). The LoD can be adjusted based on the object's distance from the viewer, object significance, or position (Biljecki et al. 2014; Abualdenien and Borrmann 2022). By incorporating LoD techniques, rendering efficiency is bolstered through a reduction in the workload on graphics pipelines. The slight decrease in visual fidelity of distant or less-significant objects is often imperceptible due to its minimal impact on object appearance (Clark 1976). In recent years, LoD techniques have garnered significant interest in the domain of the neural radiance field. There are two primary pathways for realizing objects with varying LoD: transitioning from low detail to high detail and vice versa (Heok and Daman 2004). Noteworthy works in enhancing detail from low levels include BungeeNeRF (Xiangli et al. 2022), Mip-NeRF (Barron et al. 2021), and LoD-NeuS (Zhuang et al. 2023). BungeeNeRF enhances NeRF by progressively activating high-frequency channels in NeRF's positional encoding inputs, gradually unveiling more intricate details as training progresses. Notable works in reducing detail from high to low include NGLoD (Takikawa et al. 2021), which represents implicit surfaces using an octree-based feature volume that adeptly fits shapes with multiple discrete LoDs. With advancements in 3DGS techniques, researchers have started exploring modeling different LoD in explicit 3D Gaussian scenes (Kerbl et al. 2024). For instance, Fischer et al. (2024) integrated 3D Gaussians as an efficient geometry scaffold while utilizing neural fields as a compact and flexible appearance model. CityGaussian (Liu et al. 2024) generates varying detail levels through the progressive compression strategy LightGaussian (Fan et al. 2023), operating directly on trained Gaussians to achieve substantial compression rates with minimal performance degradation.

In contrast to methods that transition from high-detail models to low-detail models in static Gaussian models, our work focuses on generating avatars that progresses from low-level detail to high-level detail. Specifically, for dynamic Gaussian avatars, Gaussians are embedded within the local coordinate system of triangle faces rather than the global world coordinate system. This embedding strategy results in Gaussians associated with different face indices being unable to merge, even when they are close to each other Dongye et al. (2024). A more appropriate approach for dynamic Gaussian avatars is to employ a low-to-high detail generation process. Our method enables the creation of drivable Gaussian avatars with a controllable number of Gaussians, achieving a smooth transition from low to high detail through hierarchical embedding and selective detail enhancement.

## 3 Method

### 3.1 Overview

In Sect. 3, we elaborate primarily on utilizing two methods, hierarchical embedding and selective detail enhancement, to produce Gaussian avatars characterized by high visual quality and minimal computational costs. Our methodology is delineated into four primary phases: data preparation, Gaussian embedding, Gaussian optimization, and selective detail enhancement. The data preparation phase is instrumental in generating homogeneous topological mesh avatar and training data for Gaussian embedding, as depicted in Fig. 2(a). Subsequently, the Gaussian embedding, Gaussian optimization, and selective detail enhancement phases are instrumental in crafting Gaussian avatars with varying LoD, as illustrated in Fig. 2(b). These resultant Gaussian avatars are capable of real-time driving and application in interactive VR settings, as demonstrated in Fig. 2(c). Given the advancements over traditional 3DGS offered, we initially expound upon the fundamentals of 3DGS in Sect. 3.2, followed by a detailed exposition of the four sequential phases of our methods in Sects. 3.3–3.6, respectively.

### 3.2 Preliminary

The 3DGS technique is employed to depict virtual objects or scenes using anisotropic 3D Gaussians, which are determined by the image and camera parameters. The virtual entities are represented by a 3D Gaussian series with 3D covariance matrix $\Sigma$ and mean $\mu$.

$$G(x) = e^{-\frac{1}{2}(x-\mu)^T \Sigma^{-1}(x-\mu)} \tag{1}$$

To optimize these Gaussians efficiently through gradient descent, Kerbl et al. introduce parametric ellipses by utilizing a scaling matrix $S$ and a rotation matrix $R$ to construct the covariance matrix.

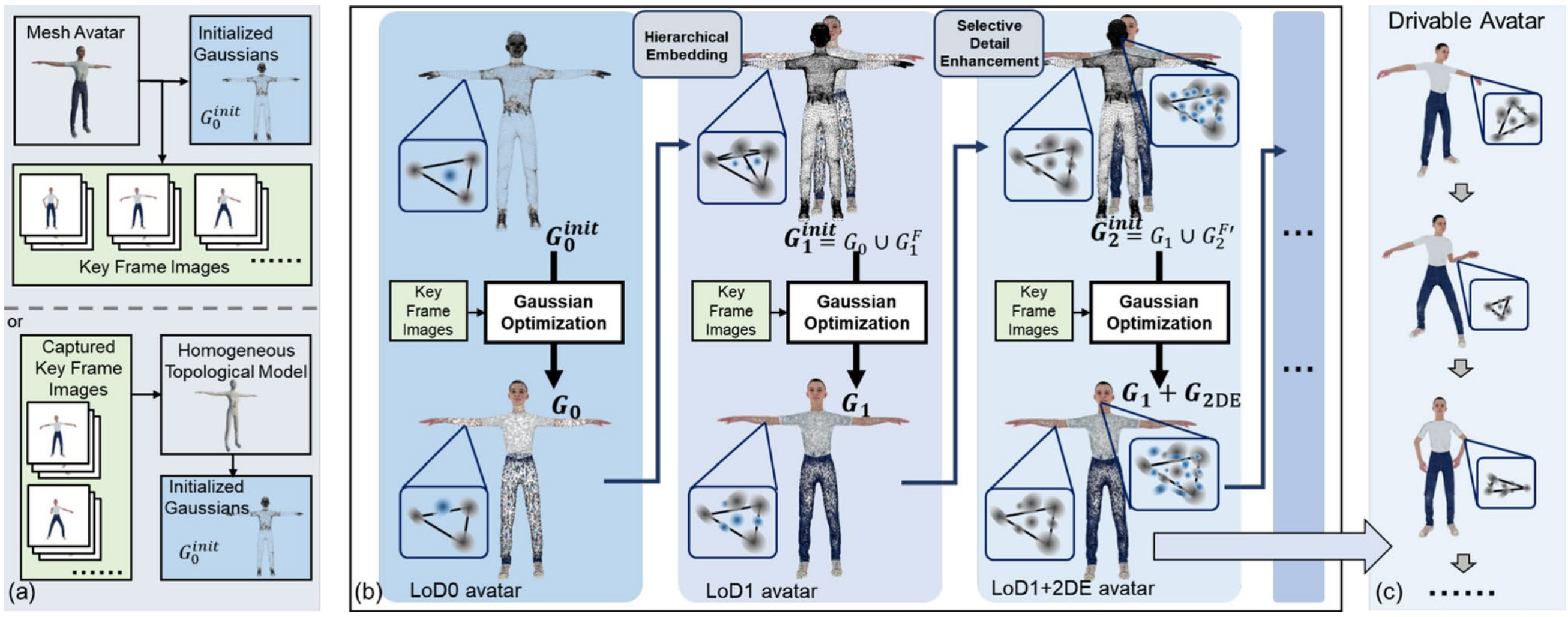

**Fig. 2** The pipeline of LoDAvatar. **a** In the data preparation phase, the homogeneous topological mesh avatar and a sequence of key frame images with the corresponding camera parameters are generated. **b** Local coordinate systems are established on the mesh's triangle faces, followed by subsequent Gaussians embedding and Gaussian optimization. Various LoDAvatars are generated via hierarchical embedding and selective detail enhancement. **c** The Gaussians are embedded in the local coordinate systems of the triangle faces, synchronizing with the movements of the mesh to create drivable Gaussian avatars

$$\Sigma = RSS^T R^T \tag{2}$$

During the rendering process, the projection of Gaussians from 3D space to a 2D image plane is executed via a view transformation $W$ and the Jacobian of the affine approximation of the projective transformation $J$. The covariance matrix $\Sigma'$ in the 2D image plane can be calculated as

$$\Sigma^{'} = JW\Sigma W^T J^T \tag{3}$$

The color $C$ of a pixel is determined by blending all overlapping 3D Gaussians with

$$C = \sum_{i=1}^{N} c_i \alpha_i \prod_{j=1}^{i-1} (1 - \alpha_j) \tag{4}$$

where $c_i$ represents the color of each point and $\alpha_i$ is derived by evaluating a 2D Gaussian with covariance $\Sigma$ multiplied by a learned per-point opacity.

In summary, as a representation of a virtual object or scene, the 3D Gaussians encompass the following parameters: (1) World Position $X \in R^3$, (2) World Rotation $r \in R^4$ expressed as quaternion, (3) Scaling Factor $s \in R^3$, (4) Spherical Harmonics Coefficients for color information $h \in R^{48}$, and 5) Opacity $\alpha \in [0, 1]$.

### 3.3 Data preparation

In the data preparation phase, we aim to prepare a homogeneous topological mesh avatar and multi-view key frame images with their corresponding camera parameters. There are two approaches for data preparation: based on an existing mesh avatar or based on captured multi-view key frame images.

The mesh avatar itself has topological consistency, allowing us to directly generate multiple key frame poses. The mesh avatars can be sourced from platforms such as Adobe (2025). We position 42 virtual cameras with identical intrinsic parameters at different positions and angles around the avatar in a spherical arrangement. The mesh avatar is placed at the center of the spherical array of cameras, with a radius of 2 meters, ensuring that the virtual cameras capture the entire avatar during the filming. Subsequently, we capture multi-view images at each key frame, setting the resolution to $1080 \times 1080$, and record the corresponding camera parameters. All keyframes will be utilized to optimize the Gaussians in future Gaussian optimization processes.

The alternative method involves using real-world captured key frame images as input. This involves aligning the images at each key frame, performing feature point matching, and generating a sparse point cloud along with a set of camera positions. A dense point cloud is then constructed based on the estimated camera positions and the images (Schönberger and Frahm 2016). Following this, the mesh is reconstructed from the dense point cloud within the Agisoft Metashape software (Agisoft 2025). Since the mesh avatars generated from different frames lack topological consistency, it is necessary to retopologize the mesh avatar of each frame to a mesh model that shares the same triangle indices. We select the Skinned Multi-Person Linear (SMPL) model (Loper et al. 2015), which has been widely adopted as a standard avatar in previous research. The SMPL model is a vertex-based 3D human model comprising 6890 vertices and 13,776 triangle faces, capable of representing various human shapes and poses through 10 shape parameters and 23 joint points. During the SMPL fitting, if the avatar's clothing in the original images closely fit to the body, the SMPL model can be automatically fitted using a multi-view SMPL fitting method (Bojanić 2024). However, if the avatar features loose or complex clothing, which prevents an accurate fitting through multi-view SMPL fitting, each frame mesh should be manually retopologized to a precise SMPL model. After retopologization, the mesh avatars share the same vertices and faces, rendering them suitable for subsequent Gaussian embedding.

### 3.4 Gaussian embedding

Gaussian embedding is conducted on the mesh avatar as outlined in Sect. 3.3. The primary objective of Gaussian embedding is to establish a linkage between the avatar represented with mesh and the avatar represented with Gaussians. During rendering, the Gaussian avatar is set to visible while the mesh avatar is set to invisible, thereby reducing computational costs during rasterization (Guédon and Lepetit 2024). Initially, the world coordinates of the vertices on each triangle face of the mesh avatar are acquired. For each triangle, a local coordinate system is established with the center position of the triangle serving as the origin. The z-axis is aligned with the normal direction of the triangle face, the x-axis points towards the first vertex, and the y-axis is determined as the cross of $z$ and $x$ directions.

Drawing on the Gaussian embedding method from Gaussian Avatars (Qian et al. 2024), we select four locations on each triangle, i.e. the three vertices and the triangle's center for Gaussian initialization, as illustrated in Fig. 3(a). The rationale for initializing Gaussians at the centers and vertices of the triangles is twofold. First, the scales of the triangles are relatively uniform in standard mesh models. Initializing Gaussians at the centers ensures a more even distribution of Gaussians across the avatar surface, which in turn accelerates the optimization process and allows for quicker convergence. Second, Gaussians are embedded in multiple iterations, with new triangles being generated at each stage. Gaussians initialized at the centers of triangles in one iteration become the Gaussians at the vertices of new triangles in the subsequent iteration. To maintain consistency across multiple embedding iterations, we initialize the Gaussians at each vertex only once, where multiple triangle faces share the same vertices.

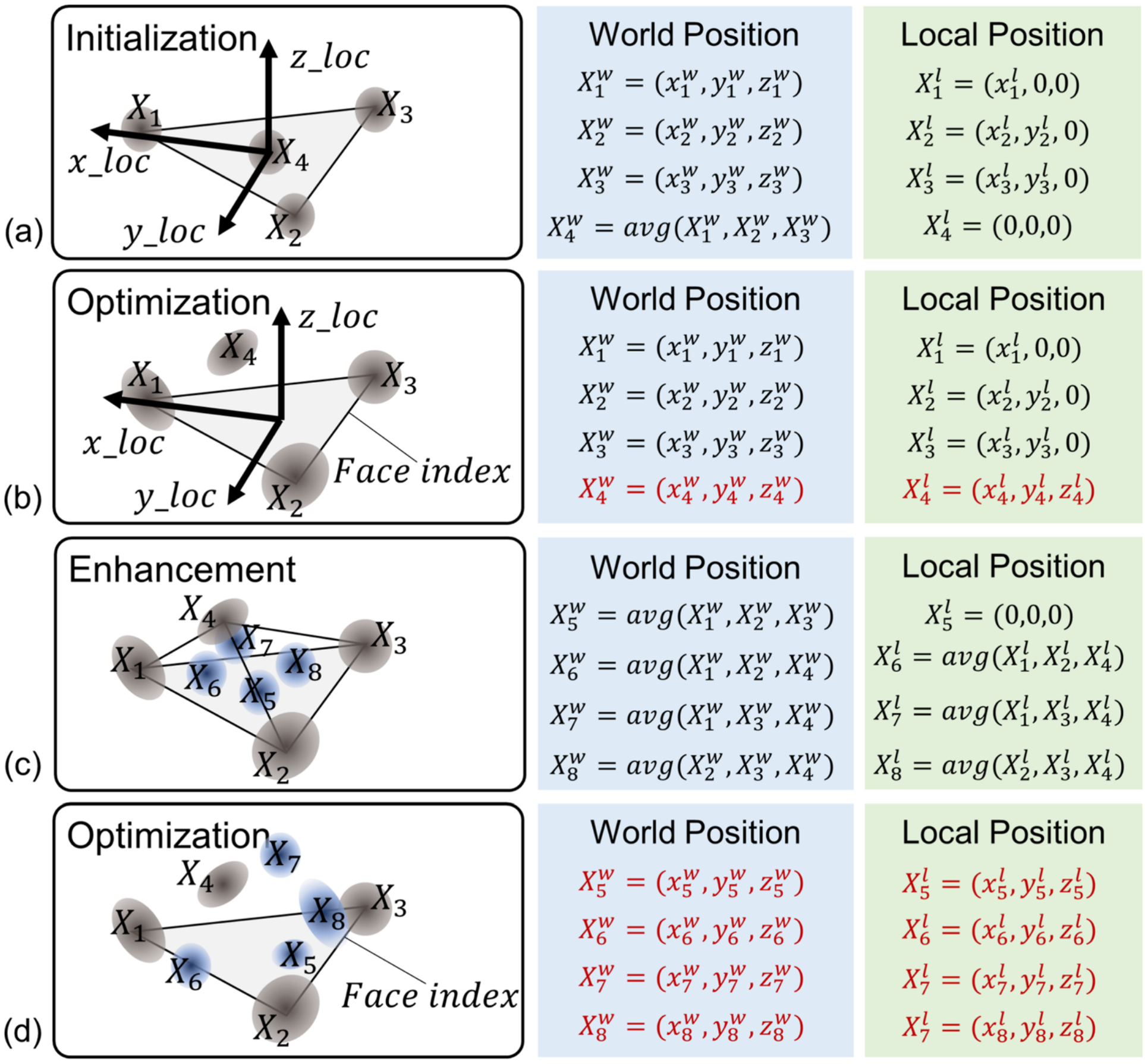

**Fig. 3** Hierarchical embedding on the triangle faces. **a** Establishing local coordinate systems and initializing Gaussians. **b** Fixing the position of the Gaussians at the vertices and performing Gaussian optimization. **c** Connecting the positions of optimized Gaussians, forming new triangle faces, and initializing new Gaussians. **d** Fixing existing Gaussians and optimizing new Gaussians to achieve enhanced LoD

Within the local coordinate system of the triangle, each Gaussian is endowed with four new variables: *idx*, $X^l$, $r^l$, and $s^l$. Here, *idx* denotes the triangle face number to which the Gaussians are embedded; $X^l$ signifies the Gaussian's local position on the triangle face; $r^l$ represents the local rotation; and $s^l$ denotes the local scale. The embedded Gaussian's parameters can be defined by:

$$G = \begin{cases} X^w = k \cdot R \cdot X^l + T \\ r^w = R \cdot r^l \\ s^w = k \cdot s^l \\ h^w = h^l \\ \alpha^w = \alpha^l \end{cases} \tag{5}$$

where *R* denotes the rotation matrix from local to the world coordinate system; *T* signifies the translation matrix; *k* represents the scaling factor for the triangle face area variation upon mesh movement; the color parameter *h* and transparency $\alpha$ remain constant. The Gaussians embedded in the triangles adjust accordingly as the mesh avatar moves. During the subsequent Gaussian optimization process, the position, rotation, and scale parameters of the Gaussians in the local coordinate system are optimized. For different poses captured by multiple key frame images, the parameters of the Gaussians in the local coordinate system are first transformed into the world coordinate system, based on the changes in the mesh triangle faces at the current position. The loss is then computed to ensure that the set of local parameters can adapt to the variations in different poses. Consequently, the mesh-Gaussian embedding on triangles is established, yielding the initialized Gaussian parameter $G_0^{init}$.

## 3.5 Gaussian optimization

The original 3DGS encompasses two primary optimization steps: Gaussian parameter optimization and adaptive density control. Our Gaussian optimization process differs from the original 3DGS in two main aspects. First, during Gaussian parameter optimization, we optimize the Gaussian parameters in the local coordinate system rather than in the world coordinate system. Second, to achieve Gaussian avatars with distinct LoD, our method omits the use of adaptive density control in the Gaussian optimization phase. Specifically, we avoid the splitting and pruning sessions to maintain a constant number of Gaussians. This decision is motivated by the aim to ensure that each Gaussian, at varying LoD, effectively retains the essential characteristics of the human avatar throughout the optimization process. The Gaussian parameters at lower LoD are held constant in each successive iteration. Adding details is incorporated into the avatar at lower LoD by introducing and optimizing new Gaussians, thereby establishing a hierarchical embedding framework for Gaussian avatars.

In this work, we adopt the loss function from the original 3DGS paper (Kerbl et al. 2023) and Gaussian Avatars (Qian et al. 2024), using $L = 0.8 \times L_1 + 0.2 \times L_{D-SSIM}$ for optimizing the position, rotation, scaling, opacity, and color coefficient. Here $L_1$ is the $L_1$ loss between the rendered image and the original image. $L_{D-SSIM}$ represents the structural similarity index measure loss between the images. We experimented with various weight combinations, starting with the 0.8 and 0.2 weights, and found that convergence was faster when retaining these weights. We extract the camera parameters and project the initialized $G_0^{init}$ onto the image planes, comparing them with the multi-view key frame images to compute the loss function. The initialized Gaussians $G_0^{init}$ is divided into two components: Gaussians at the vertices of the triangle faces $G_0^V$ and Gaussians on the triangle faces $G_0^F$, represented as

$$G_0^{init} = G_0^V \cup G_0^F = \{X_0^V, r_0^V, s_0^V, h_0^V, \alpha_0^V\} \cup \{X_0^F, r_0^F, s_0^F, h_0^F, \alpha_0^F\} \tag{6}$$

During the optimization process, we maintain the position parameter $X_0^V$ fixed, and solely optimize the parameters $r_0^V$, $s_0^V$, $h_0^V$, $\alpha_0^V$ of $G_0^V$ and $G_0^F$, as depicted in Fig. 3(b). This consideration avoids redundant storage of the Gaussian avatar at lower levels of detail, thereby diminishing memory consumption and ensuring rapid optimization. Following optimization, we obtain the Gaussian avatar $G_0$ at the lowest level of detail, denoted as

$$\begin{aligned} G_0 &= G_0^{V'} \cup G_0^{F'} \\ &= \{\mathbf{X_0^V}, r_0^{V'}, s_0^{V'}, h_0^{V'}, \alpha_0^{V'}\} \cup \{X_0^{F'}, r_0^{F'}, s_0^{F'}, h_0^{F'}, \alpha_0^{F'}\} \\ &= G_1^V \end{aligned} \tag{7}$$

When $G_0$ is driven, the Gaussians on each face adjust with their corresponding triangles.

Subsequently, additional details are incorporated into $G_0$ to achieve a Gaussian avatar with a higher LoD. Post the initial optimization, the Gaussians at the center of the triangle faces transition to new local positions $X_0^{F'}$. For each triangle in the mesh avatar, the position of each $X_0^{F'}$ is linked to the vertices of its respective triangle face, forming three new triangle faces, as illustrated in Fig. 3(c). Four new Gaussians are initialized at the center positions of the newly formed triangle faces and the original triangle face. At this stage, $G_0$ can be viewed as the Gaussian at the vertex positions, while the newly embedded Gaussians can be seen as the Gaussians on the faces, represented as

$$G_1^{init} = G_1^V \cup G_1^F = G_0 \cup \{X_1^F, r_1^F, s_1^F, h_1^F, \alpha_1^F\} \tag{8}$$

In the subsequent optimization step, we maintain the parameters of $G_0$ fixed, reintroduce the multi-view key frame images and camera parameters, and optimize solely the newly introduced $G_1^F$ to capture additional details in the avatar. This iterative process enables the gradual augmentation of details while managing the number of Gaussians through hierarchical embedding. The Gaussian avatar at a higher level of detail is denoted as

$$G_1 = G_1^V \cup G_1^{F'} = G_0 \cup \{X_1^{F'}, r_1^{F'}, s_1^{F'}, h_1^{F'}, \alpha_1^{F'}\} = G_2^V \tag{9}$$

At this juncture, five Gaussians are embedded in each triangle face of the original mesh, all adjusting with their corresponding triangle faces when manipulated, as depicted in Fig. 3(d).

To generate Gaussian avatars with increased LoD, connecting $\{X_5, X_1\}$, $\{X_5, X_2\}$, $\{X_5, X_3\}$, $\{X_6, X_1\}$, $\{X_6, X_2\}$,$\{X_6, X_4\}$,$\{X_7, X_1\}$,$\{X_7, X_3\}$,$\{X_7, X_4\}$,$\{X_8, X_2\}$, $\{X_8, X_3\}$, $\{X_8, X_4\}$ results in creating 12 new triangles, initializing the embedding of $(4 + 12 = 16)$ additional Gaussians on the surface. By repeating this process iteratively and continuously optimizing the newly generated Gaussians, we can progressively enhance the existing Gaussian avatar. In scenarios where the mesh avatar employed for embedding is an SMPL model, following 3 to 4 iterations of Gaussian embedding and optimization, the formation of $G_2$ and $G_3$ avatars, each comprising $296k - 1.17M$ Gaussians, can yield a visually superior Gaussian avatar observable from diverse viewpoints.

### 3.6 Selective detail enhancement

In Sect. 3.5, we embed Gaussians to transform a low-detail avatar into a high-detail avatar. Within each iteration, every triangle face of the original mesh avatar undergoes subdivision into four new triangle faces, accompanied by the initialization of new Gaussians on these faces. In addition, the selective detail enhancement method is introduced to regulate the division of triangle faces and the initialization of new Gaussians. The core concept behind selective detail enhancement lies in achieving targeted enhancement by selectively addressing specific triangle faces, embedding, and initializing new Gaussians while creating a more detailed Gaussian avatar. Past studies have underscored the substantial impact of the quality on avatars' faces (Suk and Laine 2023; Latoschik et al. 2017) and hands (Jiang et al. 2023) on interactions within VR (Roth et al. 2016). Hence, enhancing designated details can be accomplished by selectively implementing Gaussian embedding on the facial and manual triangles of the mesh avatar. This method offers the advantage of enabling additional control over the number of Gaussians, facilitating the acquisition of crucial details for the face and hands of the Gaussian avatar while minimizing additional computational costs. Consequently, this strategy enhances the user's visual subjective experience within the VR.

The core of the selective detail enhancement lies in identifying which triangles require further subdivision. For standard topology mesh models, this can be accomplished by specifying the triangle indices. For other mesh models, given the challenge of directly obtaining the indices of face and hand triangles on the mesh avatar, we employ image masks to identify the triangles corresponding to the avatar's head and hands. It is preferable to choose one or more images with clear poses and no occlusions and then determine the indices of the triangles that need further detail enhancement through mask segmentation and projection. For example, in the scenario illustrated in Fig. 4, we opt for a frontal view image captured when the avatar assumes a T-pose, along with the associated internal and external camera parameters, denoted as $K$ and $[R|T]$ for this specific image. Leveraging the avatar, we can derive the world coordinates $[x, y, z]$ of all vertex positions within the mesh avatar. Subsequently, employing the prevalent image segmentation algorithm SAM (Kirillov et al. 2023), we extract the masks representing the avatar's head and hands. Following this, we project all vertex positions within the mesh avatar onto the image plane based on the internal and external camera parameters corresponding to the image, denoted as $[u, v, w] = K[R|T][x, y, z, 1]^T$. This projection process assists in determining whether a vertex falls within the designated mask. Upon confirming that all three vertices of a triangle lie within the mask, we record their indices, forming the triangle face. During a specific iteration transitioning from the low-detail to high-detail avatar, these indices can be selectively utilized to generate new triangle faces with Gaussian embedding. This targeted approach aims to enhance the facial and manual areas of the Gaussian avatar, imbuing the avatar with crucial details essential for optimal performance in real-time interactive VR environments.

## 4 Evaluation

Section 4 evaluates the proposed LoDAvatar method through two experiments. LoDAvatar, a Gaussian avatar generation method based on hierarchical embedding and selective detail enhancement, offers enhanced control over the number of Gaussians compared to previous approaches. The evaluation encompasses two primary aspects. Experiment 1 assesses the ability of the method to generate high visual quality Gaussian avatars and measures the optimization time and the real-time frame rate. Experiment 2 examines frame rate fluctuations when rendering one or multiple

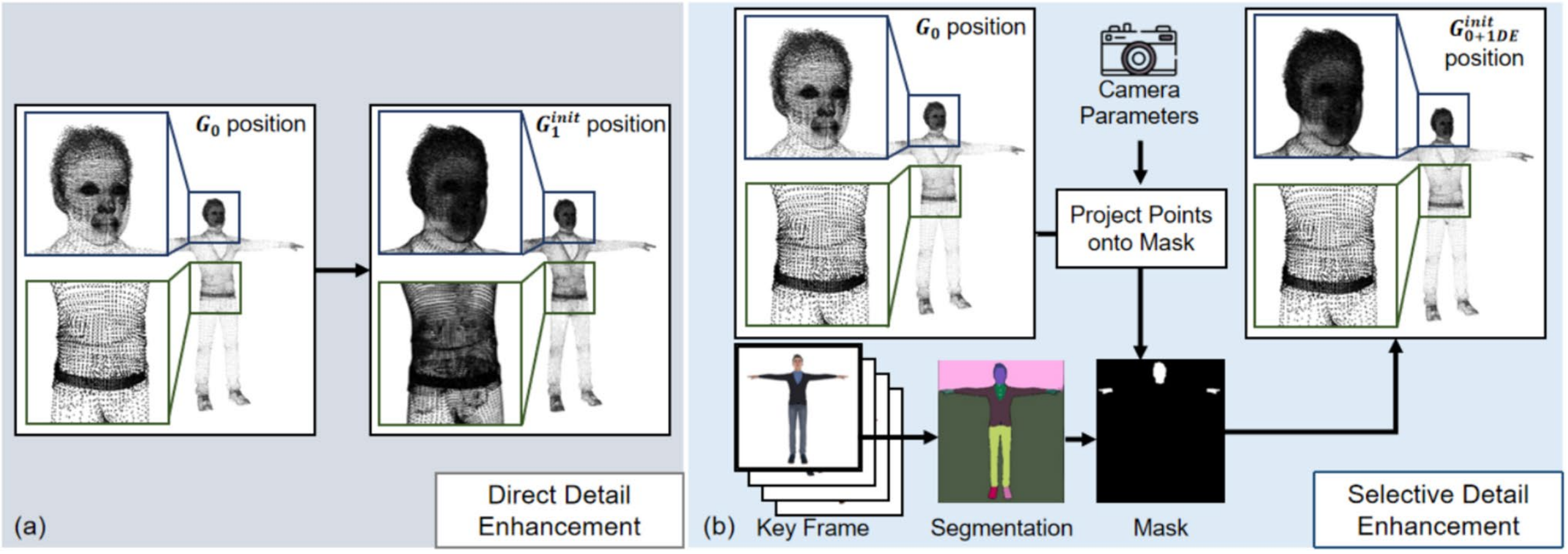

**Fig. 4** **a** Direct detail enhancement embeds new Gaussians on all triangle faces. **b** Selective detail enhancement embeds new Gaussians solely on the triangle faces selected by a mask, allowing for targeted enhancement in specific areas to regulate the number of Gaussians

Gaussian avatars simultaneously to evaluate the associated computational costs.

## 4.1 Gaussian avatars and baselines

In Section 4, our evaluation encompasses eight Gaussian avatars: (1) LoD1, (2) LoD2, (3) LoD3, (4) LoD1 with LoD2 detail enhancement (LoD1+2DE), (5) LoD1 with LoD3 detail enhancement (LoD1+3DE), (6) LoD2 with LoD3 detail enhancement (LoD2+3DE), (7) Gaussian Avatars, and (8) Splatting Avatars. These avatars are derived from the same mesh avatar, i.e. SMPL, as initial inputs within each experiment. Specifically, the LoD1, LoD2, and LoD3 are formulated through the hierarchical embedding twice, thrice, and four times, respectively, as delineated in Sects. 3.3–3.5. The multi-view key frame images used for Gaussian optimization are the same for all Gaussian avatars, regardless of their level of detail. These avatars also share the same SMPL model. The absence of the LoD0 is due to the SMPL model, which features only 6890 vertices and 13,776 faces. The LoD0 initialized with merely 20,666 Gaussians, leading to visible transparent body parts during motion. Expanding upon the LoD1 and LoD2, we establish the (4) LoD1+2DE, (5) LoD1+3DE, and (6) LoD2+3DE using the selective detail enhancement method described in Sect. 3.6.

Regarding the baselines, we chose Gaussian Avatars (Qian et al. 2024) and Splatting Avatars (Shao et al. 2024), both of which are closely related to our method and produce high-quality avatars. In Gaussian Avatars, a Gaussian is initialized at the center of each triangle face in the mesh avatar and subsequently subdivided through adaptive density control. However, during optimization, the densification and pruning processes are uncontrollable and cannot be manually intervened. This method is initially applied to facial datasets, and we tested it on full-body avatars in this experiment. Splatting Avatars is the current state-of-the-art method that initializes Gaussians randomly and embeds each Gaussian to the corresponding faces through the 'walking on triangle' during optimization. This method allows for an arbitrary number of Gaussians at initialization, with automatic pruning during optimization. While these two methods share similarities with our embedding approach, they initialize and optimize Gaussians on mesh avatars in a single step. Unlike LoDAvatar, neither Gaussian Avatars nor Splatting Avatars allows for an adjustable number of Gaussians during avatar generation. By comparing our method with these two methods, we can effectively evaluate the performance of our approach.

## 4.2 Experiment 1: visual quality—objective assessment

### 4.2.1 Dataset

In objective assessments, we employed the People Snapshot Dataset (Alldieck et al. 2018) to train the 8 different avatars. This dataset comprises standard SMPL (Loper et al. 2015) avatars, along with corresponding textures, commonly utilized in previous studies for evaluating the quality of Gaussian avatars (Wu et al. 2024). Specifically, we chose the 'female-3-casual', 'male-2-casual','female-4-sport' and 'male-3-sport'. The 'female-3-casual' and 'male-2-casual' avatars are characterized by a combination of high-frequency and low-frequency details. In contrast, the 'female-4-sport' avatar predominantly features low-frequency details with fewer high-frequency details, while the 'male-3-sport' avatar has more high-frequency details and fewer low-frequency details. Given that all mesh avatars share the same topology as SMPL, each avatar is initialized with an identical number of Gaussians, regardless of the pose. In the data preparation phase, following the method described in Sect. 3.3, we positioned 42 virtual cameras within a spherical space with a radius of 2 meters around each of these avatars. We selected 40 key frame poses from the animations, capturing 42 different perspective images for each key frame. Each image was standardized to a resolution of $1080 \times 1080$, with the corresponding camera parameters recorded. Subsequently, leveraging this dataset, we utilized this data to undergo Gaussian embedding, Gaussian optimization, and selective detail enhancement, creating 8 distinct Gaussian avatars.

#### 4.2.2 Evaluation

Building upon evaluation from existing research (Qian et al. 2024), we employed two setups to assess the visual quality of different Gaussian avatars: (1) novel-view (employing key frame poses from training sequences to animate Gaussian avatars and rendering from novel viewpoints) and (2) reenactment (animating the avatars with different poses and rendering all 42 camera views). Drawing on the objective evaluation metrics for Gaussian avatars from prior research, given the availability of dense ground truth views in our experiments (Qian et al. 2024; Yu et al. 2024), we opted for full-reference (FR) metrics instead of no-reference (NR) metrics (Mittal et al. 2012; Qu et al. 2023, 2024) to evaluate the image similarity between rendered images and ground truth (Yang et al. 2024). The FR metrics used include *PSNR*, *SSIM*, and *LPIPS*. Additionally, we measured the frames per second (FPS) when rendering a single avatar. In this evaluation, we refrained from conducting assessments of avatars at distinct LoD across varying observation distances. This choice was deliberate, as we utilized a uniform white background during Gaussian optimization, rendering image similarity metrics unsuitable for evaluating visual quality across diverse observation distances.

The optimization time for drivable Gaussian avatars depends on various factors. These factors include the avatar's intrinsic attributes (e.g. skin and clothes), texture complexity, texture resolution, numbers of keyframe images, number of cameras, and specific CPU and GPU (Peng et al. 2024). Due to the multitude of factors influencing optimization time, prior works have not established a universal evaluation method (Fei et al. 2024). Thus, we define our own method for testing the optimization time. The loss function during Gaussian optimization in the original paper (Kerbl et al. 2023) is expressed as $L = 0.8 \times L_1 + 0.2 \times L_{D-SSIM}$. After every 10,000 epochs of Gaussian optimization, we calculate the loss for the current epoch. If the loss reduction is less than $0.5\%$ after 10,000 epochs, we consider the Gaussian optimization complete and record the optimization time. In our experiments using the People Snapshot Dataset, we documented the average optimization time for eight distinct Gaussian avatars across the four models.

#### 4.2.3 Results

**Visual Quality.** In the contexts of novel-view and reenactment, Tables 1 and 2 present the image similarity metrics for the two distinct models among the eight Gaussian avatars, we highlight the top three digits of each metric in bold. The objective assessment of image similarity reveals that when the Gaussian counts are limited and the avatar is at a lower

**Table 1** Novel-view image similarity metrics

| Avatar | | LoD1 | 1+2DE | 1+3DE | LoD2 | 2+3DE | LoD3 | GA (Qian et al. 2024) | SA (Shao et al. 2024) |
|---|---|---|---|---|---|---|---|---|---|
| Female-3-casual | PSNR ↑ | 14.87 | 16.23 | 18.58 | 24.11 | **28.61** | **30.34** | 27.88 | **30.20** |
| | SSIM ↑ | 0.55 | 0.67 | 0.72 | 0.83 | 0.91 | **0.98** | **0.92** | **0.98** |
| | LPIPS ↓ | 0.31 | 0.26 | 0.24 | 0.12 | **0.06** | **0.04** | **0.06** | **0.03** |
| | FPS ↑ | **256** | **245** | **240** | 223 | 210 | 162 | 235 | 150 |
| Male-2-casual | PSNR ↑ | 13.69 | 14.22 | 16.93 | 22.83 | **27.74** | **30.14** | 25.29 | **30.99** |
| | SSIM ↑ | 0.51 | 0.65 | 0.73 | 0.91 | **0.93** | **0.96** | 0.89 | **0.97** |
| | LPIPS ↓ | 0.33 | 0.28 | 0.22 | 0.11 | **0.08** | **0.04** | 0.10 | **0.03** |
| | FPS ↑ | **253** | **248** | **237** | 225 | 209 | 158 | 224 | 152 |
| Female-4-sport | PSNR ↑ | 15.17 | 17.98 | 19.23 | 25.02 | 29.41 | **30.57** | **30.25** | **30.43** |
| | SSIM ↑ | 0.62 | 0.68 | 0.76 | 0.88 | **0.92** | **0.98** | **0.92** | **0.98** |
| | LPIPS ↓ | 0.31 | 0.25 | 0.22 | 0.12 | **0.05** | **0.04** | 0.06 | **0.04** |
| | FPS ↑ | **265** | **258** | **242** | 235 | 223 | 162 | 240 | 168 |
| Male-3-sport | PSNR ↑ | 13.67 | 14.56 | 16.23 | 22.02 | **26.83** | **29.69** | 26.52 | **29.77** |
| | SSIM ↑ | 0.50 | 0.64 | 0.70 | 0.83 | 0.85 | **0.92** | **0.88** | **0.91** |
| | LPIPS ↓ | 0.30 | 0.29 | 0.25 | 0.14 | **0.11** | **0.09** | 0.12 | **0.09** |
| | FPS ↑ | **243** | **237** | **221** | 215 | 202 | 156 | 218 | 162 |

**Table 2** Reenactment image similarity metrics

| Avatar | | LoD1 | 1+2DE | 1+3DE | LoD2 | 2+3DE | LoD3 | GA (Qian et al. 2024) | SA (Shao et al. 2024) |
|---|---|---|---|---|---|---|---|---|---|
| Female-3-casual | PSNR ↑ | 11.12 | 14.54 | 14.76 | 20.17 | **22.48** | **26.42** | 22.42 | **27.82** |
| | SSIM ↑ | 0.42 | 0.62 | 0.69 | 0.79 | **0.88** | **0.92** | 0.87 | **0.93** |
| | LPIPS ↓ | 0.38 | 0.29 | 0.26 | 0.15 | **0.08** | **0.07** | 0.10 | **0.05** |
| | FPS ↑ | **205** | **194** | **161** | 129 | 121 | 61 | 154 | 87 |
| Male-2-casual | PSNR ↑ | 10.98 | 12.87 | 14.82 | 20.13 | 23.17 | **26.13** | **23.47** | **26.12** |
| | SSIM ↑ | 0.48 | 0.59 | 0.62 | 0.78 | 0.84 | **0.93** | **0.86** | **0.93** |
| | LPIPS ↓ | 0.37 | 0.31 | 0.25 | 0.13 | 0.12 | **0.08** | **0.11** | **0.07** |
| | FPS ↑ | **203** | **198** | **157** | 131 | 119 | 58 | 140 | 92 |
| Female-4-sport | PSNR ↑ | 11.33 | 15.12 | 16.11 | 21.22 | **24.93** | **27.99** | 22.08 | **27.82** |
| | SSIM ↑ | 0.52 | 0.65 | 0.70 | 0.80 | **0.87** | **0.94** | 0.86 | **0.94** |
| | LPIPS ↓ | 0.33 | 0.28 | 0.24 | 0.13 | **0.08** | **0.06** | 0.11 | **0.06** |
| | FPS ↑ | **210** | **194** | **159** | 134 | 120 | 56 | 131 | 94 |
| Male-3-sport | PSNR ↑ | 10.25 | 12.67 | 14.56 | 19.45 | **23.12** | **25.07** | 22.98 | **25.28** |
| | SSIM ↑ | 0.49 | 0.59 | 0.61 | 0.75 | **0.84** | **0.90** | 0.83 | **0.92** |
| | LPIPS ↓ | 0.39 | 0.32 | 0.28 | 0.19 | **0.14** | **0.10** | 0.15 | **0.09** |
| | FPS ↑ | **194** | **188** | **154** | 118 | 110 | 48 | 127 | 82 |

LoD, the image similarity tends to be low due to the number of Gaussians is insufficient to represent all the details of the original mesh avatar. As the number of Gaussians increases, allowing a more complete representation and increasing the details of the avatar, higher image similarity metrics are achieved. Within the settings of novel-view and reenactment, the LoD2+3DE avatars and LoD3 avatars generated through our methods demonstrate commendable image similarity metrics, nearly reaching or surpassing those of the GA and SA groups. This observation suggests that via iterative hierarchical embedding, we can progressively generate Gaussian avatars with higher LoD from low-detail Gaussian avatars, thereby consistently enhancing their visual quality.

**FPS.** During the rendering process, Gaussian avatars require that all Gaussians be sorted and subsequently projected onto the camera's image plane, thereby incurring a time complexity of $O(nlogn)$, where $n$ represents the number of Gaussians. While low-detail level avatars display lower image similarity metrics relative to high-detail level avatars, they attain superior FPS performance in real-time rendering scenarios. This observation suggests that low-detail level avatars are not simply transitional states en route to generating high-detail level avatars. At considerable viewing distances, low-detail level avatars can effectively serve as proxies for high-detail level avatars. A more detailed evaluation for low-detail level avatars will be presented in Sect. 4.3.

**Optimization Time.** In our experiments, we employed an Intel Core i7 processor and an NVIDIA RTX 3080 graphics card. The average optimization time for LoD0 avatars was 11.6 hours. Building on LoD0 avatars, the average optimization time for LoD1 was 3.4 hours. Building on LoD1 avatars, the average optimization time for LoD1+2DE was 1.8 hours, and for LoD2, it was 4.4 hours. Building on LoD1+2DE avatars, the average optimization time for LoD1+3DE was 1.6 hours. Building on LoD2 avatars, the average optimization time for LoD2+3DE was 3.1 hours, and for LoD3, it was 8.9 hours. For baselines, the average optimization time for GA was 23.4 hours, and for SA, it was 32.0 hours. Unlike GA and SA, which perform Gaussian embedding and continuous densification and pruning in a single optimization step, our method restricts the number and position of Gaussians during the Gaussian embedding and optimization phases, resulting in faster optimization times. For the high visual quality avatar LoD2+3DE, the total average generation time was $11.6 + 3.4 + 4.4 + 3.1 = 22.5$ hours, which is less than the average time required by the GA and SA groups. In LoDAvatar, the Gaussian embedding and optimization for high-detail avatars are based on low-detail avatars. The high-detail avatars share Gaussian parameters with the low-detail avatars, thus reducing the optimization time.

**Dataset.** In the People Snapshot Dataset, avatars with predominantly low-frequency details exhibit higher image similarity metrics compared to those with a greater number of high-frequency details. We attribute this to the fact that a limited number of Gaussians, which constitute the avatar, are insufficient to accurately represent all high-frequency information, such as complex patterns and wrinkles on clothing. When the original data contains more high-frequency details, selecting avatars with higher detail levels can yield better visual quality.

### 4.3 Experiment 1: visual quality—subjective evaluation

#### 4.3.1 Dataset and setup

In the subjective evaluation, we aim to assess participants' subjective perception of avatars' visual quality across varying viewing distances. Diverging from objective assessments, we did not utilize the People Snapshot dataset due to its limited texture resolution of $256 \times 256$. In a VR context, where users engage with 3D avatars while immersed in a head-mounted display (HMD), higher-resolution textures are imperative for 3D avatars compared to evaluating 2D avatars on a screen (Slater 2018). During the data preparation phase for subjective evaluation, we employed 42 real cameras with a resolution of $4096 \times 4096$ to capture real human data for 40 key frames. We performed photo-alignment on the multi-view images of each key frame using Metashape software and generated mesh avatars from the height field. The post-alignment images were segmented to obtain white background images, and we recorded the corresponding camera parameters for each image. Following this, we manually retopologized the mesh avatar of each key frame in the SMPL model. The subsequent generation of Gaussian avatars was consistent with the method described in Sect. 4.2, all derived from the same SMPL model with shared topological structures, with the segmented images and camera parameters.

We utilized the mesh avatars generated by Metashape as the baseline for the subjective evaluation, with each mesh avatar consisting of approximately 1,100,000 triangle faces. We opted not to use the retopologized SMPL model as the baseline due to its insufficient number of triangle faces, which resulted in notably poorer visual quality. The participants could easily discern the triangles in VR. For evaluation, participants were randomly presented with eight Gaussian avatars and one mesh avatar at various distances, with the order of presentation randomly selected. Our subjective evaluation comprised both static and dynamic avatar assessments. In the static setting, the avatars maintained a fixed T-pose standing position, while in the dynamic setting, the avatars played a 30-second animation. The different Gaussian avatars were deployed within the Unity Gaussian Splatting project (Pranckevicius 2024). The Gaussian position point cloud maps corresponding to the eight Gaussian avatars and schematic diagrams illustrating observations at different distances are shown in Fig. 5.

#### 4.3.2 Participants and experiment procedure

A total of 15 participants took part in the subjective evaluation, comprising 6 females and 9 males, aged between 22 and 36 years (M = 27.2). Eight of them had prior experience with VR devices. The testing was set up using Unity 2022.3.24f1. Each evaluation session featured only one type of avatar and background within the scene. The inclusion of the background helped the participants perceive the current observation distance. The participants used Meta Quest 3 HMD and conducted the experiments while seated. The avatars appeared directly in front of the participants at horizontal distances of 1 m, 5 m, and 9 m, corresponding to the distances of interaction near, medium and far in the human-avatar interaction (Rogers et al. 2022). The eight Gaussian avatars and a baseline mesh avatar were presented to the participants in random order. Furthermore, the observation distances between participants and avatars were randomized, with different types of avatars appearing at varying observation distances. After observing each avatar for a minimum of 30 s, participants were tasked with rating the visual quality of the avatars they observed using a 7-point Likert Scale. Participants evaluated each avatar until they rated all static and dynamic avatars at different observation distances and all rating data were recorded for analysis.

#### 4.3.3 Results

The subjective evaluations of the Gaussian avatars at various LoD and baselines when observed from different distances in both static and dynamic scenarios are depicted in Fig. 6. Each participant completed the entire evaluation and the scores from 15 participants exhibited a normal distribution. The comparisons include four main aspects: 1. The evaluation of the same avatar at different distances reveals differences in subjective visual quality, supporting the feasibility of introducing LoD into dynamic Gaussian avatars. 2. Comparing different avatars at the same distance further demonstrates that LoDAvatar can adapt to various observation distances. 3. The assessment of selective detail enhancement confirms its effectiveness. 4. Comparisons with the GA and SA groups show that LoDAvatar achieves higher overall subjective visual quality.

1. **Comparing the same avatar at different observation distances.** The data was analyzed using the Repeated Measures ANOVA analysis. In the static setting, significant differences were found for LoD1, LoD2 and LoD3 avatars ($F_{(2,42)} = 131.20$, $p<0.001$). In the dynamic setting, significant differences were also found for LoD1, LoD2 and LoD3 avatars ($F_{(2,42)} = 192.96$, $p<0.001$). The results suggest that high LoD avatars demonstrate consistent subjective visual quality across multiple observation distances, whereas low LoD avatars exhibit differences. Consequently, the avatar's LoD

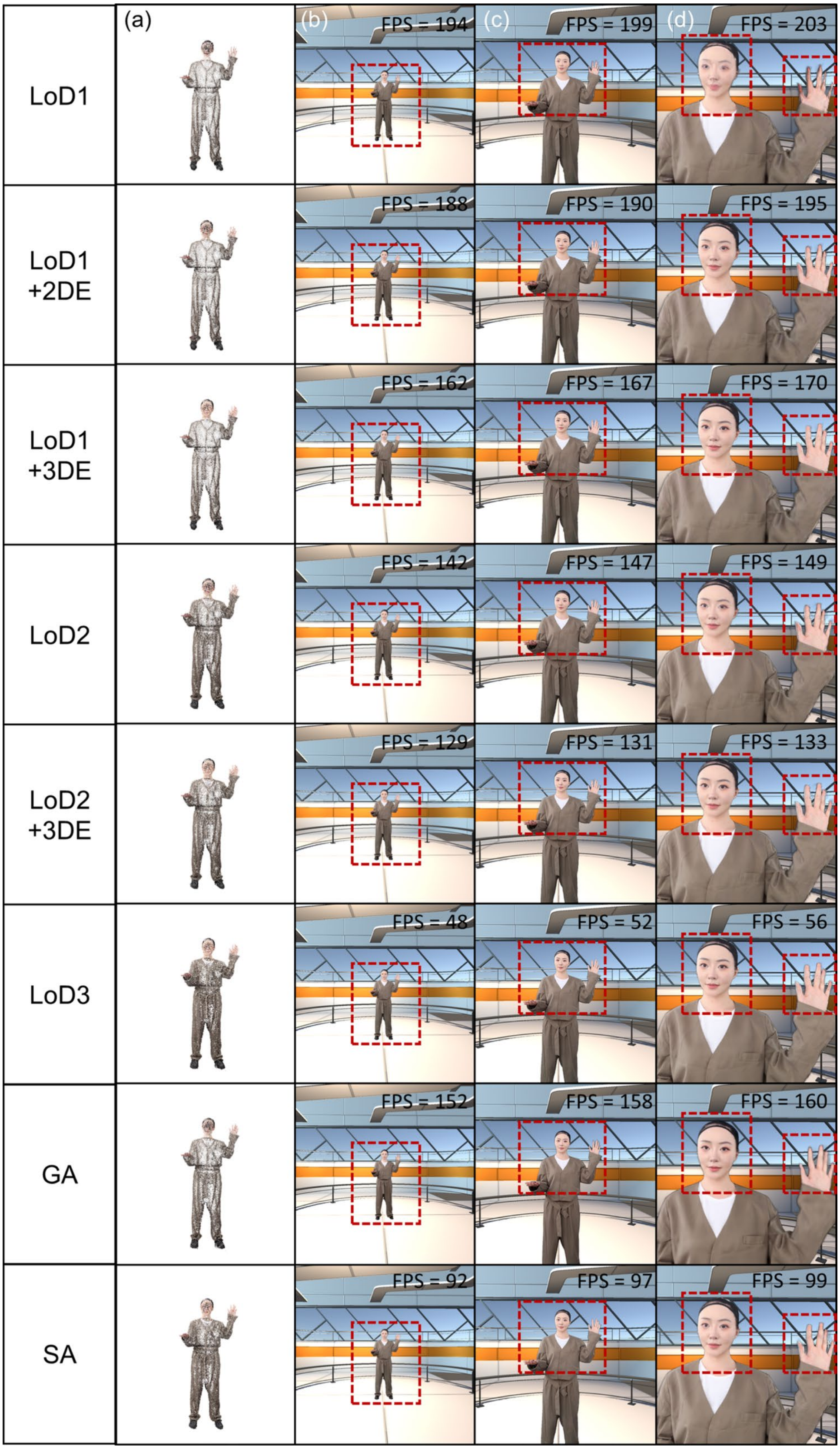

**Fig. 5** Gaussian avatars in the subjective evaluation. **a** The Gaussian point cloud positions. **b**–**d** Gaussian avatars and background observed at 9 m, 5 m and 1 m distances

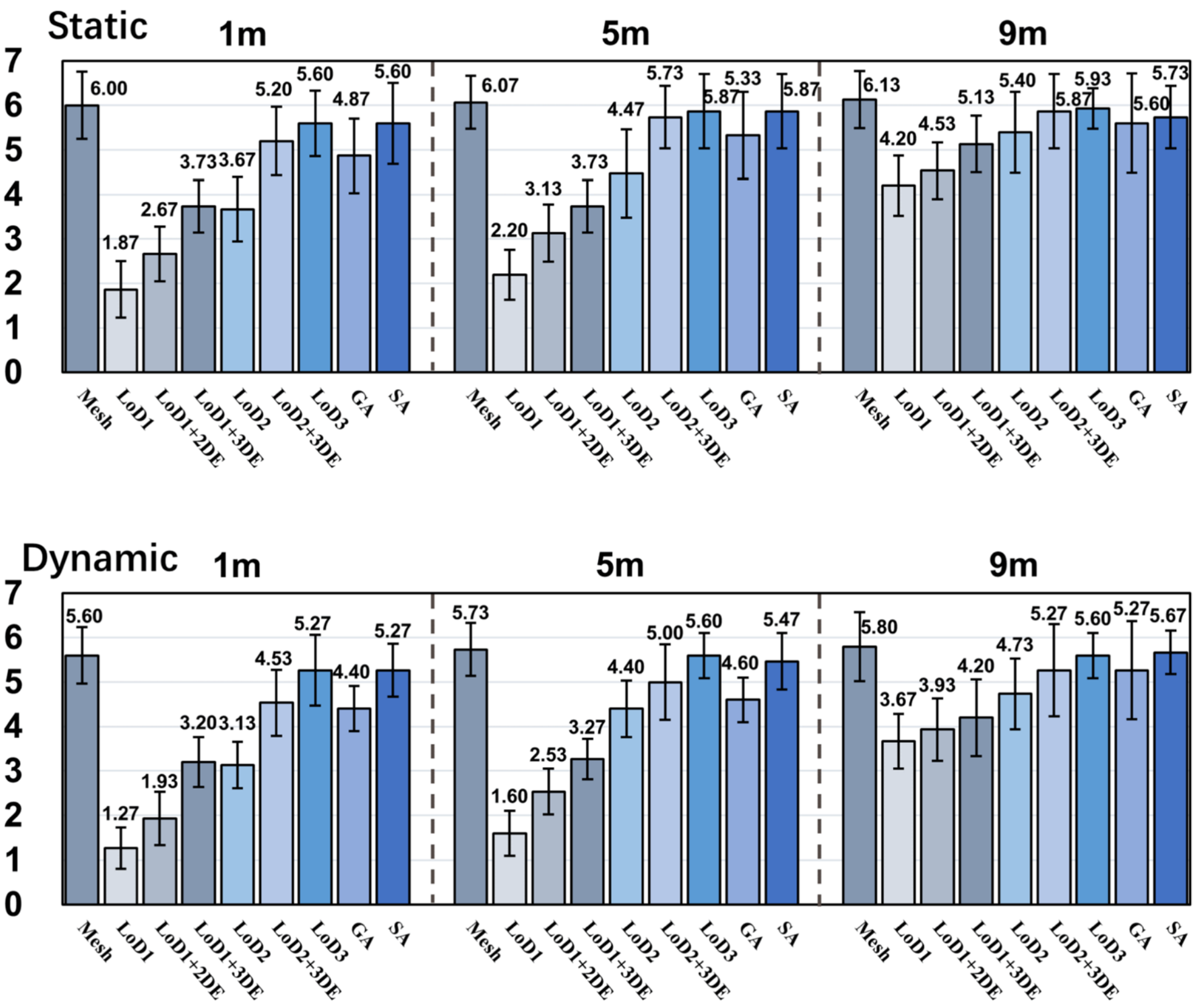

**Fig. 6** The subjective score of mesh avatar and different Gaussian avatars in static and dynamic settings. The error bar indicates the standard deviation

can be dynamically adjusted based on the human-avatar distance to mitigate visual quality variations.

2. **Comparing different avatars at the same distance.** The data was analyzed using paired t-tests. At a $1m$ observation distance, significant differences were observed between LoD1 and LoD2 (static(S) $t_{14} = -9.00$, $p<0.001$); dynamic(D) $t_{14} = -14.00$, $p<0.001$); significant differences were observed between LoD2 and LoD3 (S: $t_{14} = -7.25$, $p<0.001$; D: $t_{14} = -6.59$, $p<0.001$). At a $5m$ observation distance, significant differences were observed between LoD1 and LoD2 (S: $t_{14} = -6.58$, $p<0.001$; D: $t_{14} = -11.52$, $p<0.001$); significant differences were observed between LoD2 and LoD3 (S: $t_{14} = -3.61$, $p=0.003$; D: $t_{14} = -6.87$, $p<0.001$). At a $9m$ observation distance, significant differences were observed between LoD1 and LoD2 (S: $t_{14} = -4.29$, $p=0.001$; D: $t_{14} = -9.03$, $p<0.001$); significant differences were observed between LoD2 and LoD3 (S: $t_{14} = -2.26$, $p=0.041$; D: $t_{14} = -4.52$, $p<0.001$). As the human-avatar distance increases, significant differences between low LoD and high LoD avatars persist. However, the decrease in mean scores and t-values suggests a reduction in these discrepancies. This implies that with an increasing observation distance, low LoD Gaussian avatars can still maintain a certain level of visual quality and receive subjective evaluations that progressively approach those of high LoD avatars.
3. **Comparing with/without selective detail enhancement.** The data was analyzed using paired t-tests. At a $1m$ observation distance, significant differences were observed between LoD1 and LoD1+2DE (S: $t_{14} = -4.58$, $p<0.001$; D: $t_{14} = -3.57$, $p=0.003$); significant differences were also found between LoD1 and LoD1+3DE (S: $t_{14} = -7.90$, $p<0.001$; D: $t_{14} = -10.64$, $p<0.001$); there were significant differences between LoD2 and LoD2+3DE (S: $t_{14} = -5.28$, $p<0.001$; D: $t_{14} = -8.57$, $p<0.001$). At a $5m$ observation distance, significant differences were observed between LoD1 and LoD1+2DE (S: $t_{14} = -4.09$, $p=0.001$; D: $t_{14} = -6.09$, $p<0.001$); significant differences were also found between LoD1 and LoD1+3DE (S: $t_{14} = -9.28$, $p<0.001$; D: $t_{14} = -7.17$, $p<0.001$); significant differences were observed between LoD2 and LoD2+3DE (S: $t_{14} = -4.46$, $p=0.001$; D: $t_{14} = -3.15$, $p=0.007$). At a $9m$ observation distance, there were no significant differences between static LoD1 and LoD1+2DE ($t_{14} = -1.78$, $p=0.096$), but significant were found in dynamic LoD1 and LoD1+2DE ($t_{14} = -2.26$, $p=0.041$). there were significant differences between LoD1 and LoD1+3DE (S: $t_{14} = -4.09$, $p=0.001$; D: $t_{14} = -3.23$, $p=0.006$); significant differences were also observed between LoD2 and LoD2+3DE (S: $t_{14} = -2.17$, $p = 0.048$; D: $t_{14} = -4.00$, $p = 0.001$). Furthermore, at $1m$ distance, there were no significant differences between LoD2+3DE and LoD3 (S: $t_{14} = -1.47$, $p = 0.164$; D: $t_{14} = -2.05$, $p = 0.060$); at $5m$ distance, there were no significant differences between LoD2+3DE and LoD3 ($t_{14} = -0.69$, $p = 0.499$), but there was a significant difference dynamically ($t_{14} = -3.15$, $p = 0.007$); at $9m$ distance, there were no significant differences between LoD2+3DE and LoD3 (S: $t_{14} = -0.367$, $p = 0.719$; D: $t_{14} = -1.58$, $p = 0.136$). These results indicate that through facial and manual detail enhancement, avatars can enhance subjective visual quality evaluations while only slightly increasing the number of Gaussians, thus achieving a balance between visual quality and computational costs. These results highlight the importance of selective detail enhancement.

4. **Comparing with baseline and existing methods.** The LoD2+3DE received favorable evaluations in both static and dynamic at all distances. We conducted paired t-tests to compare LoD2+3DE with the baseline mesh avatar, GA, and SA. At a $1m$ observation distance, there were no significant differences between LoD2+3DE and GA (S: $t_{14} = 1.16$, $p = 0.265$; D: $t_{14} = 0.62$, $p = 0.546$); there were no significant differences between LoD2+3DE and SA (S: $t_{14} = -1.47$, $p = 0.164$; D: $t_{14} = -1.78$, $p = 0.096$); significant differences were observed between LoD2+3DE and the baseline (S: $t_{14} = -2.45$, $p = 0.028$; D: $t_{14} = -4.68$, $p < 0.001$). At a $5m$ observation distance, there were no significant differences between LoD2+3DE and GA in the static settings ($t_{14} = 1.57$, $p = 0.138$), but significance was found in the dynamic settings ($t_{14} = 2.45, p = 0.028$); no significant differences were found between LoD2+3DE and SA (S: $t_{14} = -0.52$, $p = 0.610$; D: $t_{14} = -1.71$, $p = 0.110$); significant differences were observed between LoD2+3DE and the baseline (S: $t_{14} = -2.45$, $p = 0.028$; D: $t_{14} = -3.56$, $p = 0.003$). At $9m$ observation distance, there were no significant differences between LoD2+3DE and GA (S: $t_{14} = 0.94$, $p = 0.364$; D: $t_{14} = 0.69$, $p = 0.499$); there were no significant differences between LoD2+3DE and SA (S: $t_{14} = 0.435$, $p = 0.670$; D: $t_{14} = -1.19$, $p = 0.253$); no significant differences were found between LoD2+3DE and the baseline (S: $t_{14} = -1.00$, $p = 0.334$; D: $t_{14} = -1.95$, $p = 0.072$). The comparison results between the baseline and the LoD2+3DE suggest that although variances still exist between LoD2+3DE and the baseline at close and medium observation distances, the LoD2+3DE avatar can attain outcomes akin to the mesh avatar at far observation distances. Additionally, the LoD2+3DE avatar can produce visual quality comparable to the top existing Gaussian avatars while utilizing fewer Gaussians for avatar generation.

In summary, the subjective experiments have led to the following results:

(1) Gaussian avatars at different LoD show differences in subjective visual quality evaluations. Nevertheless, with increasing observation distance, lower LoD Gaussian avatars can maintain some visual quality, approaching subjective evaluations similar to those of higher LoDAvatars.
(2) The application of selective detail enhancement, which includes adding specific details to the face and hands, results in enhanced subjective evaluations.

## 4.4 Experiment 2: computational costs

Real-time responsiveness in human-agent interactions is paramount, necessitating reduced computational costs during rendering to ensure that avatars can sustain higher frame rates throughout VR engagements (Dongye et al. 2025). Experiment 2 primarily assesses the average frame rates of LoDAvatars when operating within VR, to evaluate the computational costs.

### 4.4.1 Avatars

In this experiment, we utilize the Gaussian avatars and the mesh avatar in Sect. 4.3 as input. Gaussian splatting diverges from conventional rendering pipelines by splatting each Gaussian onto the camera's image plane. Prior research has illustrated that when rendering the same virtual entities, avatars constructed based on Gaussians exhibit higher frame rates during runtime compared to mesh avatars, with computational costs directly correlated to the number of Gaussians (Bao et al. 2024). Among the varied Gaussian avatars generated from SMPL model, the number of Gaussians is as follows: LoD1 consists of 75,770 Gaussians, LoD1+2DE consists of 96,470 Gaussians, LoD1+3DE consists of 179,270 Gaussians, LoD2 consists of 296,186 Gaussians, LoD2+3DE consists of 378,986 Gaussians, LoD3 consists of 1,177,850 Gaussians, GA consists of 182,391 Gaussians, and SA consists of 412,338 Gaussians. The decision to select four iterations to achieve the LoD3 avatar is primarily driven by engineering considerations. The LoD3 avatar, with its substantial number of Gaussians, operates near the minimum frame rate achievable on current consumer-grade hardware. Higher detail levels would cause noticeable lag on these devices.

#### 4.4.2 Apparatus and materials

We employed a computer equipped with an Intel Core i7 processor and an NVIDIA RTX 3080 graphics card for frame rate assessments. The testing environment was set up using Unity 2022.3.24f1, with no extraneous objects present in the scene, solely Gaussian avatars. Individual tests were conducted with different numbers of avatars, encompassing static and dynamic avatar scenarios. In each scenario, the resolution of the camera within Unity was set to $3840 \times 2160$, ensuring that all avatars were captured. Each scene ran for one minute, during which we recorded the average frame rates exhibited by these avatars within Unity.

#### 4.4.3 Results

The results depicted in Fig. 7 illustrate the fluctuation in running frame rates of different Gaussian avatars, both static and dynamic, as the number of avatars increases. Increasing the avatar's LoD will also increase the computational costs as the number of Gaussians increases. Notably, for dynamic avatars, the frame rates during rendering lag behind those of static avatars. Furthermore, as the number of Gaussians rises, the decline in frame rates becomes more pronounced, highlighting the challenge posed by increased computational costs. While the mesh avatar, LoD3 avatar and SA showcased superior visual quality in Experiment 1, these avatars exhibited higher computational costs. Conversely, avatars with lower detail levels were demonstrated to achieve higher frame rates during rendering. Moreover, employing the selective detail enhancement method to enhance the head and hands of avatars with more interactive functionalities can further control the number of Gaussians, thereby reducing computational costs to a certain extent.

In summary, as the LoD of avatars increases and the quantity of avatars increases, a higher number of Gaussians results in increased computational costs. Due to the necessity of sorting the Gaussians before performing splatting during the rasterization process, computational costs increase significantly as the number of Gaussians increases. Thus, we advocate for integrating LoD into Gaussian avatars to align with real-time operational requisites for observing or engaging with avatars in VR. Designers can use Gaussian avatars with different LoD while maintaining visual quality in VR.

## 5 Discussion

### 5.1 Further discussion

As a novel rendering technique, Gaussian Splatting presents advantages in both high visual quality and low computational costs. The drivable 3D Gaussian avatar method has gained significant attention due to the need for real-time interactions with avatars in VR. Prior research has primarily concentrated on implementing drivable Gaussian avatars without extensively addressing the balance between visual quality and computational costs. As the number of Gaussians

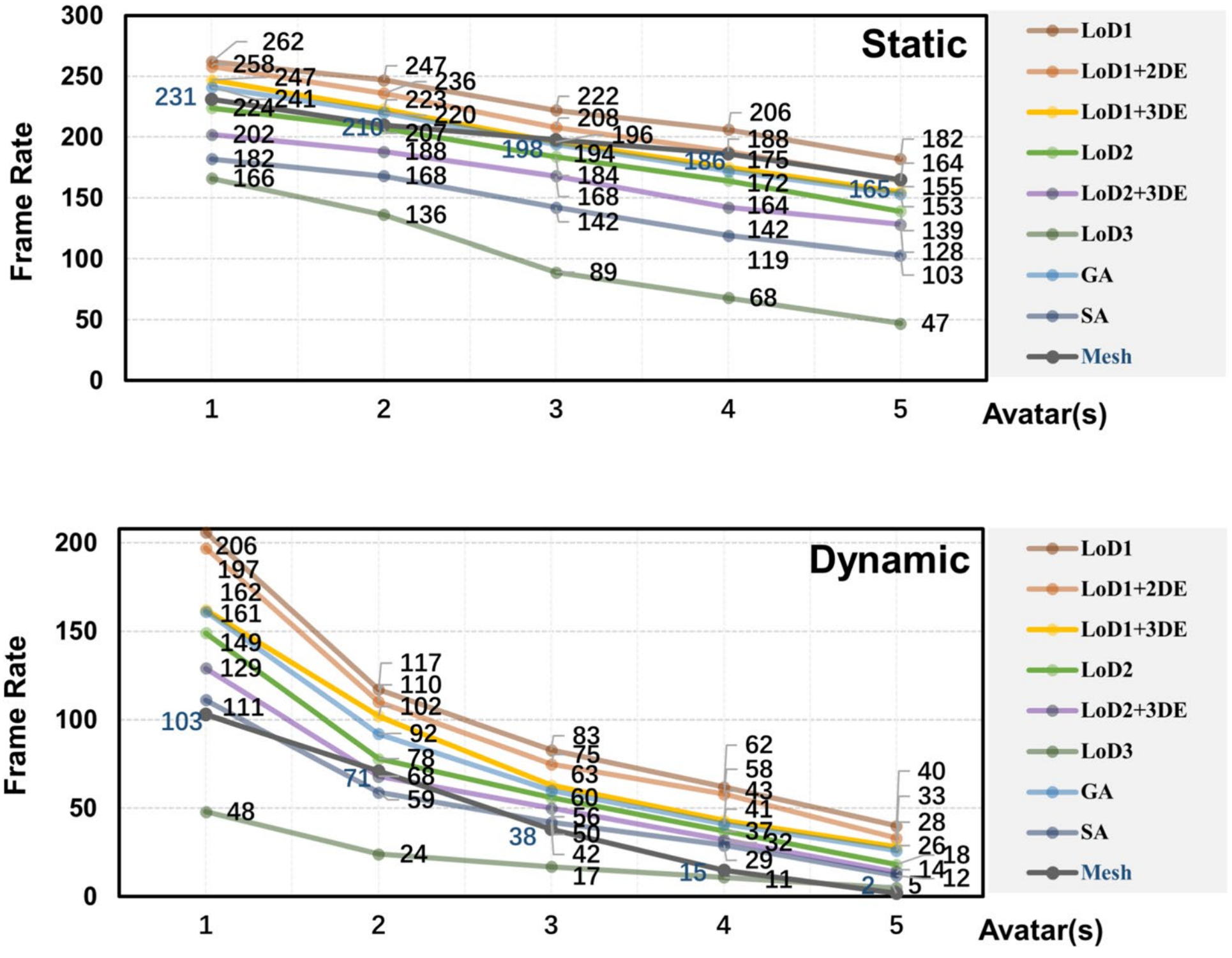

**Fig. 7** The average frame rate variation with the number of avatars for different Gaussian avatars and the mesh avatar in static and dynamic settings

increases, avatars exhibit enhanced visual quality but incur higher computational costs, resulting in a trade-off.

We present LoDAvatar, which introduces levels of detail into Gaussian avatars to better leverage the advantages of Gaussian Splatting. The methods of LoDAvatar involve hierarchical embedding and selective detail enhancement. Hierarchical embedding generates various LoD Gaussian avatars from low to high detail levels while controlling the number of Gaussians, yielding a range of avatars with diverse LoDs. Selective detail enhancement reinforces specific areas of avatars with strong interactive attributes, such as the head and hands, further controlling the number of Gaussians to reduce computational costs while minimizing the impact on visual quality. The core consideration of these two methods is to maintain a controllable number of Gaussians during avatar generation. Previous methods that employed adaptive density control resulted in an uncontrollable number of Gaussians. LoDAvatar replaces adaptive density control with hierarchical embedding. As demonstrated in Experiments 1 and 2, this substitution enables LoDAvatar to preserve visual quality while decreasing computational costs.

Employing these methods, we created avatars with varying LoD starting from a mesh avatar, and assessed the visual quality of these avatars alongside existing methods through objective and subjective evaluations. Additionally, we conducted runtime frame rate tests on different LoD avatars and existing methods to evaluate computational costs during rendering. In terms of visual quality, objective and subjective evaluations indicate that our proposed methods can increase the number of Gaussians, resulting in the generation of diverse LoDAvatars. The high LoD avatars can achieve a visual quality comparable to that of established high-quality Gaussian avatar generation methods. With an increase in the distance between participants and avatars, the subjective evaluation gap between low LoD and high LoD avatars decreases, indicating the feasibility of incorporating LoD concepts in Gaussian avatars. Furthermore, the frame rate tests for various Gaussian avatars in Experiment 2 reveal that the computational costs of rendering Gaussian avatars are linked to the number of Gaussians, with a more pronounced frame rate decrease as the number of Gaussians rises, especially noticeable in dynamic avatars.

Compared to high-quality mesh avatars, LoDAvatars maintain a similar visual quality while significantly enhancing the frame rate during rendering, particularly in scenarios where multiple avatars are present simultaneously. Additionally, the outcomes of the static avatar experiments suggest the potential for generating different LoD Gaussian objects and scenes. Our proposed method holds promise for future deployment of Gaussian avatars on mobile devices, suitable for real-time rendering of multiple Gaussian avatars or extensive Gaussian scenes. We advocate for the adoption of LoDAvatar in dynamic Gaussian avatars to strike a balance between visual quality and computational costs.

## 5.2 Limitations and future work

In this study, we introduced LoD for dynamic Gaussian avatars. However, our current work has certain limitations that suggest promising avenues for future research.

Rather than employing images directly, our method utilizes a homogeneous topological mesh avatar as input. This is because both the hierarchical embedding and selective detail enhancement stages rely on a mesh avatar with consistent topology. A mesh avatar with uniform topology ensures the consistency of Gaussian local parameters across different avatar poses and allows selective detail enhancement to be performed on triangles with the same indices. Future research could explore the automated creation of LoDAvatars directly from captured images for efficient generation. Furthermore, we implemented LoD in Gaussian avatars without initially generating distinct LoD mesh avatars for embedding. This is due to the intensive workload involved in texture resetting and skeleton re-binding on mesh avatars. Future research could investigate the visual quality disparities using Gaussian embedding on various LoD mesh avatars and hierarchical Gaussian embedding.

In our experiments, we utilized the SMPL model as the base mesh avatar. When the number of Gaussians is reduced to achieve higher performance, the avatars may exhibit blurriness or artifacts, particularly around active joints. This is due to the insufficient coverage of the model by a limited number of Gaussians, leading to transparency and visual imperfections. While the selective detail enhancement method has shown some success in mitigating blur and artifacts in the avatar's face and hands, its application to other critical joints warrants further investigation. Future work could explore advanced optimization techniques and adaptive strategies to minimize blur and artifacts, thereby expanding the applicability of our approach to a broader range of scenarios and requirements.

Within LoDAvatar, all Gaussians are embedded in local triangle face coordinate systems, with all Gaussians moving with the mesh. Future research could explore driving different components of Gaussian avatars, such as achieving more realistic movement of hair and clothing during avatar driving.

In LoDAvatar, our emphasis is solely on implementing LoD on Gaussian avatars by adjusting the position, rotation, and scale of Gaussians. Subsequent research could delve into modifying color parameters and transparency on dynamic Gaussian avatars to create a color-variable LoDAvatar for consistent lighting between avatars and backgrounds.

**Table 3** Reenactment image similarity metrics on Female-3-Casual

| Female-3-casual | | LoD1 | 1+2DE | 1+3DE | LoD2 | 2+3DE | LoD3 |
|---|---|---|---|---|---|---|---|
| T-Pose | PSNR ↑ | 5.75 | 7.26 | 8.98 | 10.12 | 12.33 | 14.05 |
| | SSIM ↑ | 0.21 | 0.24 | 0.35 | 0.47 | 0.56 | 0.66 |
| | LPIPS ↓ | 0.65 | 0.52 | 0.46 | 0.41 | 0.39 | 0.36 |
| 10 Poses | PSNR ↑ | 9.23 | 10.04 | 10.33 | 15.76 | 17.04 | 20.08 |
| | SSIM ↑ | 0.33 | 0.46 | 0.51 | 0.64 | 0.72 | 0.79 |
| | LPIPS ↓ | 0.41 | 0.33 | 0.30 | 0.20 | 0.17 | 0.15 |
| 20 Poses | PSNR ↑ | 10.42 | 12.33 | 12.45 | 18.16 | 20.07 | 23.24 |
| | SSIM ↑ | 0.39 | 0.58 | 0.60 | 0.74 | 0.83 | 0.89 |
| | LPIPS ↓ | 0.39 | 0.31 | 0.29 | 0.18 | 0.12 | 0.10 |
| 40 Poses | PSNR ↑ | 11.12 | 14.54 | 14.76 | 20.17 | 22.48 | 26.42 |
| | SSIM ↑ | 0.42 | 0.62 | 0.69 | 0.79 | 0.88 | 0.92 |
| | LPIPS ↓ | 0.38 | 0.29 | 0.26 | 0.15 | 0.08 | 0.07 |

**Table 4** Reenactment image similarity metrics on Male-2-Casual

| Male- 2 - casual | | LoD1 | 1+2DE | 1+3DE | LoD2 | 2+3DE | LoD3 |
|---|---|---|---|---|---|---|---|
| T-Pose | PSNR ↑ | 5.84 | 7.07 | 8.67 | 10.02 | 12.57 | 13.85 |
| | SSIM ↑ | 0.24 | 0.26 | 0.36 | 0.42 | 0.54 | 0.67 |
| | LPIPS ↓ | 0.62 | 0.52 | 0.43 | 0.40 | 0.33 | 0.28 |
| 10 Poses | PSNR ↑ | 9.03 | 10.22 | 10.46 | 14.97 | 16.99 | 19.27 |
| | SSIM ↑ | 0.32 | 0.44 | 0.50 | 0.65 | 0.71 | 0.80 |
| | LPIPS ↓ | 0.47 | 0.42 | 0.41 | 0.25 | 0.21 | 0.18 |
| 20 Poses | PSNR ↑ | 10.21 | 11.66 | 13.03 | 18.24 | 21.02 | 23.47 |
| | SSIM ↑ | 0.42 | 0.54 | 0.59 | 0.73 | 0.79 | 0.88 |
| | LPIPS ↓ | 0.43 | 0.39 | 0.30 | 0.22 | 0.18 | 0.14 |
| 40 Poses | PSNR ↑ | 10.98 | 12.87 | 14.82 | 20.13 | 23.17 | 26.13 |
| | SSIM ↑ | 0.48 | 0.59 | 0.62 | 0.78 | 0.84 | 0.93 |
| | LPIPS ↓ | 0.37 | 0.31 | 0.25 | 0.13 | 0.12 | 0.08 |

## 6 Conclusion

In this paper, we introduce LoDAvatar, a method utilizing hierarchical embedding and selective detail enhancement to generate Gaussian avatars with different LoD. Our method takes existing mesh avatars as input, involving data preparation, Gaussian embedding, Gaussian optimization, and selective detail enhancement. We conduct two experiments to evaluate our proposed methods. Experiment 1 assesses the visual quality through both objective and subjective evaluations, demonstrating that our proposed methods can produce LoDAvatars with commendable visual quality. In Experiment 2, we examine the average frame rates of LoDAvatars during runtime to analyze computational costs, further emphasizing the importance of integrating LoD in Gaussian avatars. LoDAvatar showcases the potential to reduce the computational costs required for rendering when observing avatars from a distance. We suggest that our methods can be effectively employed for LoD generation in dynamic Gaussian avatars, striking a balance between visual quality and computational efficiency.

## Poses for Gaussian optimization

In the visual quality objective assessment, we utilized all multi-view key frame images for Gaussian optimization, following the approach in our baselines Gaussian Avatars (Qian et al. 2024) and Splatting Avatars (Shao et al. 2024). When all key frame images are used to optimize the local position, rotation, and scale of a single set of Gaussians, these Gaussian parameters in the local coordinate system can better adapt to different mesh deformations, resulting in improved visual quality during reenactment of various poses. However, the optimal number of key frame poses for Gaussian optimization in LoDAvatar remains an open question.

To further investigate this, we conducted additional experiments in the appendix, using the same setup as in Sect. 4.2. We selected the 'female-3-casual' and 'male-2-casual' datasets for this investigation. For keyframe images, we used the official SMPL animation `mosh_cmu_0511_f_lbs_10_207_0_v1.0.2.fbx` as the training set for Gaussian optimization and `mosh_cmu_7516_f_lbs_10_207_0_v1.0.2.fbx` as the test set for reenactment. The training set comprises 601 key frame images. We selected 1 frame (T-pose), one frame every 60 frames (10 poses), one frame every 30 frames (20 poses), and one frame every 15 frames (40 poses). During Gaussian optimization, we adjusted the local parameters of the Gaussians using the T-pose, 10 poses, 20 poses, and 40 poses multi-view key frame images, respectively, while keeping all other settings identical to those in Sect. 4.2. After Gaussian optimization with different numbers of key frame images, we saved six sets of Gaussian avatars with varying LoDs. The test set includes 301 key frames, from which we selected 20 poses at intervals of 15 frames. We performed reenactment of these 20 poses using the different LoD avatars optimized with varying numbers of key frame images and assessed the average image similarity of the 20 poses. The results are shown in Tables 3 and 4.

From the visual similarity assessment metrics, it is evident that using only the T-Pose for fitting LoDAvatar results in poor visual quality during reenactment. This is because the Gaussians are optimized under a single pose condition in the local coordinate system. When the mesh avatar deforms, these Gaussian parameters fail to meet the observational needs of multiple poses. As the number of introduced key frame images increases, the visual quality of LoDAvatar during reenactment also improves. When all 40 pose key frame images are used for Gaussian optimization, LoDAvatar achieves good visual quality during reenactment. We posit that optimizing with a greater number of keyframe images yields Gaussian parameters that are more adaptable to diverse poses, thereby enhancing generalization across different mesh deformations. Additionally, we observed that the improvement in image similarity metrics when using 40 poses compared to 20 poses is marginal. We attribute this phenomenon to the fact that, under the current settings, using only 20 poses can to some extent meet the needs of the reenactment task, although optimizing with all 40 poses yields a model with better visual quality. However, it is important to note that there is no strict rule or limitation on the number of poses for Gaussian optimization. This number should be determined based on the actual numbers of the input mesh avatar's vertices and triangle, the number of keyframe poses provided in dataset, and the specific application requirements. For most cases, we still recommend using as many different poses from the dataset as possible for Gaussian optimization to enhance the generalization capability of local Gaussian parameters across various poses.

**Supplementary Information** The online version contains supplementary material available at https://doi.org/10.1007/s10055-025-01249-3.

**Acknowledgements** We would like to express our gratitude to Zhichao Li for his valuable contributions to the re-topology of the mesh avatars used in our experiment, as well as for his support in the production of the manuscript's introduction video.

**Author contributions** Author contribution: Conceptualization, X.D. and D.W.; methodology, X.D. and H.G.; software, X.D., H.G. and L.L.; validation, X.D., J.G. and L.L.; formal analysis, X.D., J.G. and H.J.; investigation, X.D., Z.T. and Y.B.; resources, D.W. and L.L.; data curation, Y.B. and Z.T.; writing—original draft preparation, X.D.; writing—review and editing, X.D., D.W., H.J. and Y.B.; visualization, X.D., D.W. and Y.B.; supervision, D.W. and H.J.; project administration, D.W., J.G. and Y.B.; funding acquisition, D.W. All authors reviewed the manuscript.

**Funding** This work was supported by the National Key R&D Program of China (No.2022YFF0902303) and the 2022 major science and technology project "Yuelu Multimodal Graph-Text-Sound-Semantic Gesture Big Model Research and Demonstration Application" in Changsha (No.kh2301019) and the Beijing Chaoyang Internet 3.0 Key Core Technology R&D Project and the Fundamental Research Funds for the Central Universities (No.2024CX06023).

**Data availability** No datasets were generated or analysed during the current study.

## Declarations

**Conflict of interest** The authors declare that they have no conflict of interest.

**Ethical approval and consent to participate** The Research Ethics Committee of Beijing Institute of Technology determined that ethical review was not required for this study, as the experimental procedures were deemed to pose no short-term or long-term risks to the participants.